\documentclass[traditabstract]{aa} %
\usepackage{txfonts}
\usepackage[pdftex]{graphicx}
\usepackage{natbib}

\newcommand{\nsvs}{NSVS1425}
\newcommand{\hs}{HS0705}

\newcommand{\msun}{$M_\odot$}
\newcommand{\mjup}{$M_\mathrm{Jup}$}

\begin{document}
\title{The quest for companions to post-common envelope binaries}
\subtitle{II. NSVS14256825 and  HS0705+6700}

\author{
Beuermann, K. \inst{1} \and 
Breitenstein, P. \inst{2}  \and 
D\d{e}bski, B. \inst{3}  \and 
Diese, J. \inst{4} \and 
Dubovsky, P.~A. \inst{5}  \and 
Dreizler, S. \inst{1} \and 
Hessman, F.~V. \inst{1} \and 
Hornoch, K. \inst{6}  \and 
Husser, T.-O. \inst{1}  \and 
Pojmanski, G. \inst {7} \and 
Wolf, M. \inst{8}  \and 
Wo\'zniak, P.~R. \inst {9} \and 
Zasche, P. \inst{8}  \and 
Denk, B. \inst {2} \and 
Langer, M. \inst {2} \and 
Wagner, C. \inst {2} \and 
Wahrenberg, D. \inst {2} \and 
Bollmann, T. \inst{4} \and
Habermann, F.~N. \inst{4} \and 
Haustovich, N. \inst{4} \and 
Lauser, M. \inst{4} \and 
Liebing, F. \inst{4} \and 
Niederstadt, F. \inst{4}
} 

\institute{
Institut f\"ur Astrophysik, Georg-August-Universit\"at, Friedrich-Hund-Platz 1, D-37077 G\"ottingen, Germany \and
Westfalen-Kolleg, Rheinische Stra{\ss}e 67, D-44137 Dortmund, Germany \and
Astronomical Observatory, Jagiellonian University, ul. Orla 171, 30-244 Krak\'ow, Poland \and
Max-Planck-Gymnasium, Theaterplatz 10,D-37073 G\"ottingen, Germany \and
Vihorlat Observatory, Mierova 4, Humenne, Slovakia \and
Academy of Sciences, Fri\v{c}ova 298, CZ-251\,65 Ond\v{r}ejov, Czech Republic \and
Warsaw University Astronomical Observatory, Al. Ujazdowskie 4, 00-478 Warszawa, Poland \and
Astronomical Institute, Faculty of Math. and Physics, Charles University, CZ-180~00 Praha 8, V~Hole\v{s}ovi\v{c}k\'ach 2, Czech Republic \and
Los Alamos National Laboratory, MS D466, Los Alamos, NM 87545, USA
}
\date{Received 16 September 2011; accepted 19 December 2011}

\authorrunning{K. Beuermann et al.} 
\titlerunning{The quest for companions to post-common envelope binaries II}

\abstract{We report new mid-eclipse times of the two close binaries
  NSVS14256825 and HS0705+6700, harboring an sdB primary and a low-mass
  main-sequence secondary. Both objects display clear variations in 
  the measured orbital period, which can be explained by the action of
  a third object orbiting the binary. If this interpretation is
  correct, the third object in NSVS14256825 is a giant planet with a
  mass of roughly 12\,\mjup. For HS0705+6700, we provide evidence that
  strengthens the case for the suggested periodic nature of the eclipse
  time variation and reduces the uncertainties in the parameters of the brown dwarf
  implied by that model. The derived period is 8.4\,yr and the mass
  is 31\,\mjup, if the orbit is coplanar with the binary. This
  research is part of the \emph{PlanetFinders} project, an ongoing
  collaboration between professional astronomers and student groups at
  high schools.}

\keywords{ Stars: binaries: close -- Stars: binaries: eclipsing --
  Stars: subdwarfs -- Stars: individual: NSVS14256825, HS0705+6700 --
  Planets and satellites: detection } 

\maketitle


\section{Introduction}

\vspace*{-1mm} Some eclipsing close binaries exhibit $O\!-\!C$
variations in the observed mid-eclipse times relative to those
calculated from an underlying linear ephemeris. An early example is
the cataclysmic variable UX~UMa, which displayed what looked like a
periodic variation with an amplitude of 150\,s and a period of
29\,yr. This modulation was interpreted as being possibly caused by a
third body \citep{natherrobinson74} or by apsidal motion
\citep{africanowilson76} until \citet{rubensteinetal91} proved its
aperiodic nature.  Since then an explanation of these variations by
processes internal to the binary has been favored, which need not be
periodic. One such process \citep{applegate92} involves changes in the
internal constitution of the secondary, which give rise to changes in
its spin and variations in the orbital period by spin-orbit coupling. 
With more data becoming available, it was realized, however, that
internal processes might be too feeble to explain the observations
\citep[e.g.][]{brinkworthetal06,chen09,schwarzetal09,potteretal11},
but this notion was opposed by \citet{wittenmyeretal11}. It was also
realized that the detached post-common envelope binaries (PCEBs)
represent more suitable laboratories for studying eclipse time
variations than the accreting cataclysmic variables, in particular,
the variety to which UX~UMa belongs. With this in mind, the
interpretation of eclipse time variations in terms of the light-travel
time effect produced by a third body was revived in a series of papers
by \citet[][and references therein]{qianetal09,qianetal10a,qianetal10b}.  
An intriguing case is the semi-detached cataclysmic variable HU~Aqr,
which displays multi-periodic $O-C$ variations. \citet{qianetal11}
attributed these variations to the effects of two planets orbiting the
binary, but \citet{horneretal11} and \citet{wittenmyeretal11}
demonstrated the secular instability of the proposed planetary system
and questioned its existence. While this case remains unresolved, the
evidence of systematic and possibly periodic eclipse time variations
in detached PCEBs increases.

The 14.7 mag star HS\,0705+6700 (in short \hs) was discovered by
\citet{drechseletal01} to be an eclipsing close binary with an orbital
period of 2.3\,h, consisting of an sdB primary and an M-dwarf
secondary. \citet{qianetal09,qianetal10a} summarized all the then
available eclipse times and, fitting a sinusoidal to the data,
suggested the presence of a brown dwarf orbiting the binary with a
period of 7.15 yr.

NSVS\,14256825 (in short \nsvs) was discovered as a 13.2 mag variable
star in the Northern Sky Variability Survey \citep{wozniaketal04} and
identified as an eclipsing binary with a 2.6\,h orbital period and an
sdB primary by \citet{wilsetal07}. The source seems to be in many
respects a twin of \hs\ and the prototype sdB/M-dwarf binary HW\,Vir
\citep{leeetal09}.

\section{Observations and data analysis}

\begin{figure*}[t]
\includegraphics[bb=61 70 538 542,height=62mm,angle=-90,clip]{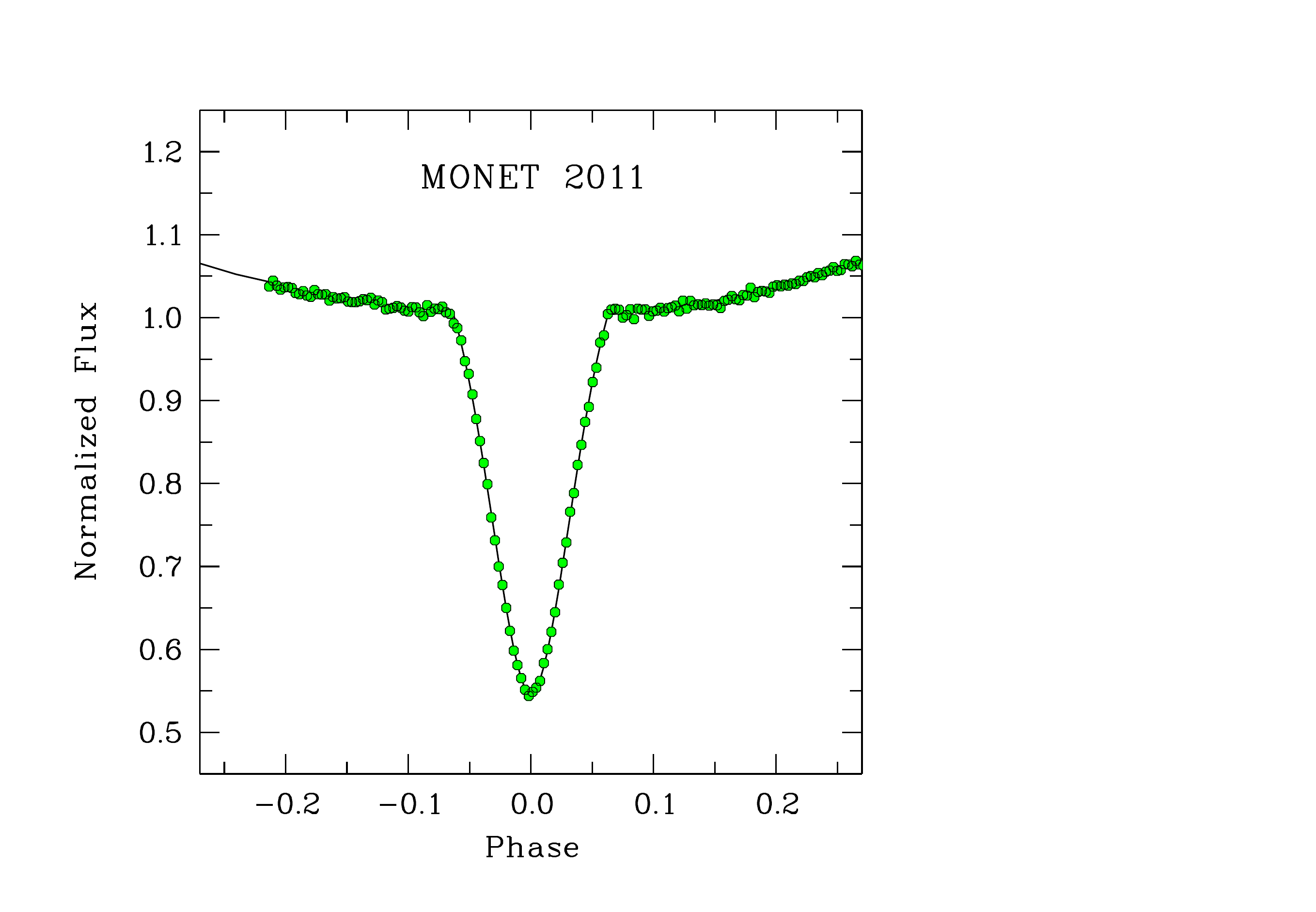}
\hspace{3mm}
\includegraphics[bb=61 107 538 542,height=57mm,angle=-90,clip]{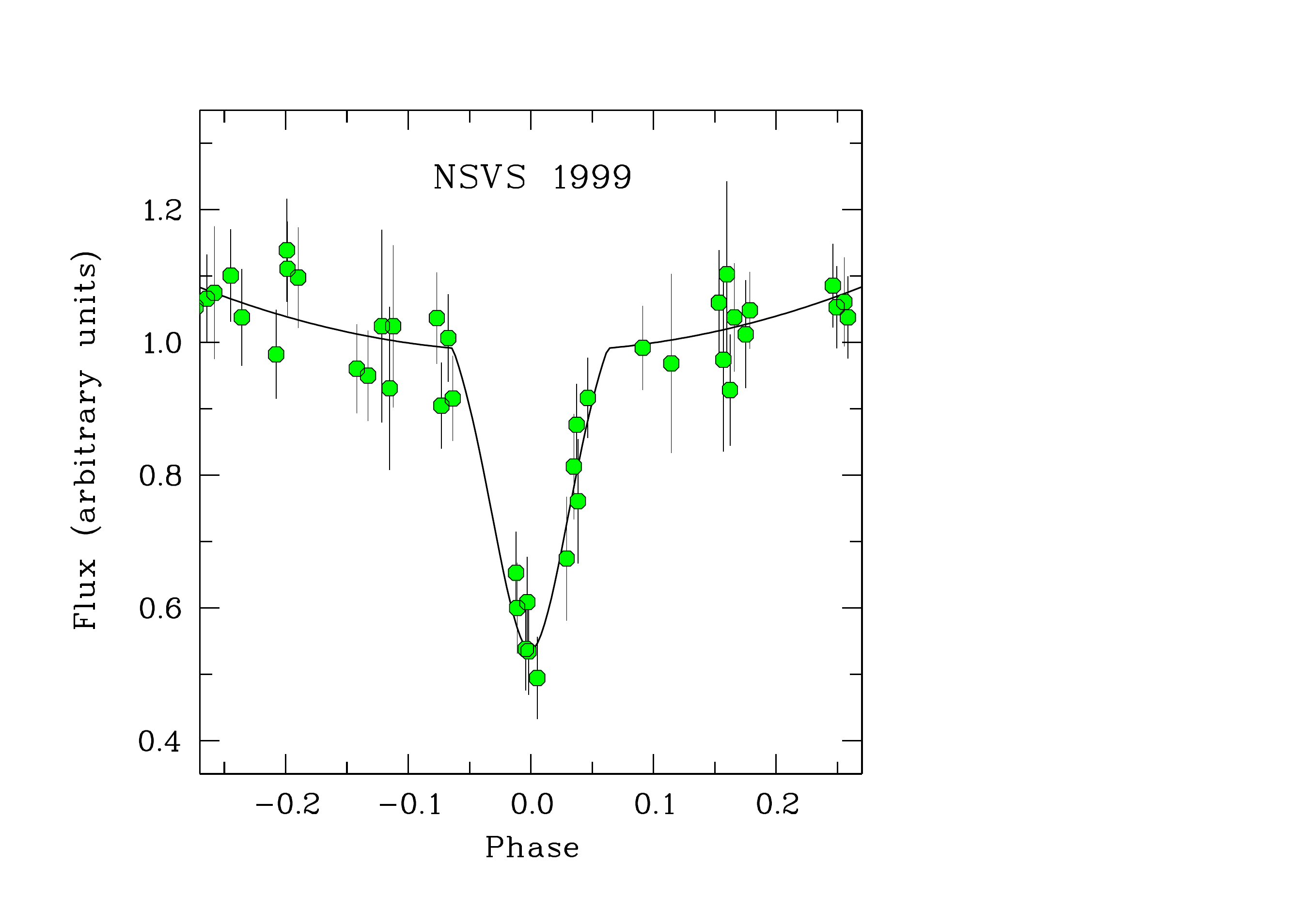}
\hspace{3mm}
\includegraphics[bb=61 110 538 542,height=57mm,angle=-90,clip]{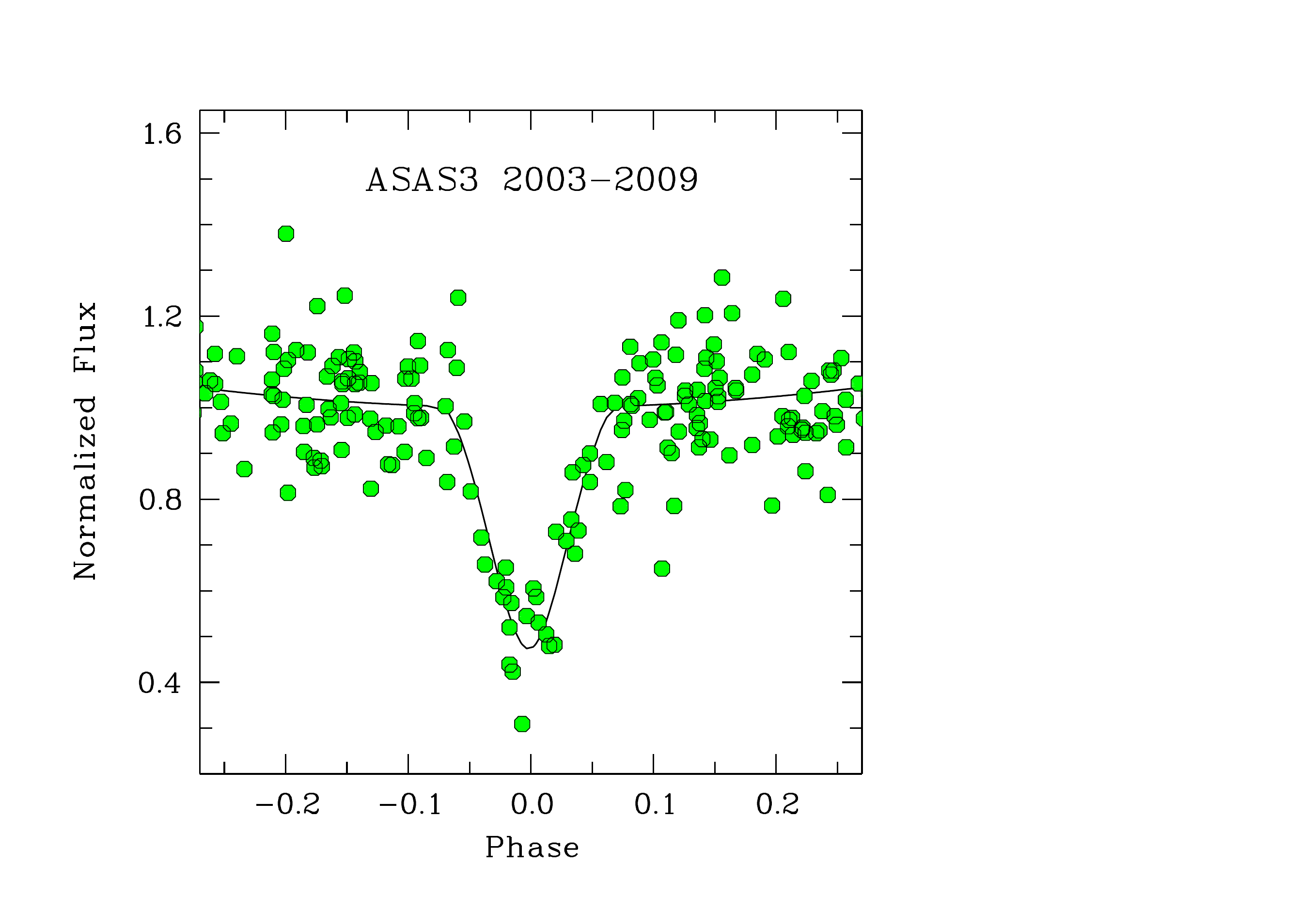}
\caption[chart]{Eclipse light curves of NSVS\,14256825. \emph{Left:}
  MONET/N on 5 May 2011, \emph{center:} Northern Sky Variability
  Survey, April -- August 1999, \emph{right:} All Sky Automated Survey 3}
\label{fig:fig1}
\end{figure*}

\begin{table}[t]
\begin{flushleft}
\caption{New mid-eclipse times of NSVS14256825  and  $O-C_\mathrm{ell}$ residuals of the elliptic orbit fit displayed
    in Fig.~2, right panel.}
\begin{tabular}{cccccc}
\hline \\[-1ex]
Cycle E   & Ecl & BJD(TT)      & Error    &  $O\!-\!C_\mathrm{ell}$  & Tel$^\dagger$ \\
       &     & (days)       & (days)   &  (days) & \\[0.5ex]
\hline\\[-1ex]
 \hspace{-3.5mm}$-$26586.0  &  I  &  51339.803273  &   0.000429  &   0.000016  & 1  \\
 \hspace{-3.0mm}$-$12390.0  &  I  &  52906.673899  &   0.000541  &   \hspace{-2.0mm}$-$0.000271  & 2 \\
 \hspace{-2.0mm}$-$5931.0   &  I  &  53619.579776  &   0.000537  &  \hspace{-2.0mm}$-$0.000573  & 2 \\
    1018.0  &  I  &  54386.569297  &   0.000569  &  \hspace{-2.0mm}$-$0.000375  & 2  \\
    3737.0  &  I  &  54686.676900  &   0.000477  &   0.000140  & 2  \\
    6914.0  &  I  &  55037.335341  &   0.000018  &  \hspace{-2.0mm}$-$0.000006  & 3 \\
    7037.0  &  I  &  55050.911367  &   0.000022  &  \hspace{-2.0mm}$-$0.000002  & 4 \\
    7304.0  &  I  &  55080.381278  &   0.000071  &   0.000008  & 3 \\
    7322.0  &  I  &  55082.368000  &   0.000021  &  \hspace{-2.0mm}$-$0.000005  & 3 \\
    9823.5  &  II &  55358.468971  &   0.000046  &  \hspace{-2.0mm}$-$0.000053  & 5 \\ 
    9959.0  &  I  &  55373.424740  &   0.000068  &   0.000012  & 5 \\
   10131.0  &  I  &  55392.409097  &   0.000018  &   0.000008  & 5 \\ 
   12763.0  &  I  &  55682.913998  &   0.000013  &   0.000013  & 6  \\
   12799.0  &  I  &  55686.887451  &   0.000013  &  \hspace{-2.0mm}$-$0.000005  & 6  \\
   12799.5  &  II &  55686.942699  &   0.000031  &   0.000056  & 6 \\
   13469.0  &  I  &  55760.838180  &   0.000027  &   0.000008  & 6 \\
   13469.5  &  II &  55760.893396  &   0.000031  &   0.000037  & 6 \\
   13470.0  &  I  &  55760.948549  &   0.000011  &   0.000003  & 6 \\
   13488.0  &  I  &  55762.935315  &   0.000018  &   0.000034  & 6 \\
   13542.0  &  I  &  55768.895489  &   0.000037  &   0.000001  & 6 \\
   13632.0  &  I  &  55778.829154  &   0.000010  &  \hspace{-2.0mm}$-$0.000012  & 6 \\
   13511.0  &  I  &  55765.473867  &   0.000017  &  \hspace{-2.0mm}$-$0.000021  & 3 \\
   13682.0  &  I  &  55784.347812  &   0.000064  &  \hspace{-2.0mm}$-$0.000064  & 5\\
   13768.0  &  I  &  55793.840061  &   0.000012  &   0.000004  & 6 \\
   13827.0  &  I  &  55800.352168  &   0.000014  &   0.000032 & 3 \\
   13828.0  &  I  &  55800.462510  &   0.000013  &   0.000000 & 3 \\
   13845.0  &  I  &  55802.338869  &   0.000024  &  \hspace{-2.0mm}$-$0.000002 & 3 \\
   13846.0  &  I  &  55802.449197  &   0.000036  &  \hspace{-2.0mm}$-$0.000048 & 3 \\
   13872.0  &  I  &  55805.318959  &   0.000015  &  \hspace{-2.0mm}$-$0.000015 & 3 \\
   13873.0  &  I  &  55805.429322  &   0.000017  &  \hspace{-2.0mm}$-$0.000026 & 3 \\
   13899.0  &  I  &  55808.299065  &   0.000016  &  \hspace{-2.0mm}$-$0.000013 & 3 \\
   14400.0  &  I  &  55863.596559  &   0.000010  &   0.000006 & 6 \\[0.5ex]
\hline\\[-2ex]                                                              
\end{tabular}
\label{tab:data1}
  
$^\dagger$ (1) NSVS, (2) ASAS3, (3) Kolonica Observatory, (4) San
Pedro Matir Observatory, (5) Ond\v{r}ejov Observatory, (6) MONET/North
\end{flushleft}
\vspace{-4mm}
\end{table}

\begin{table}[t]
\begin{flushleft}
\caption{New mid-eclipse times of HS\,0705+6700  and $O\!-\!C_\mathrm{ell}$ residuals of the elliptic orbit fit displayed in Fig.~3.}
\begin{tabular}{cccccc}
\hline \\[-1ex]
Cycle E   & Ecl & BJD(TT)      & Error    & $O\!-\!C_\mathrm{ell}$  & Tel$^\dagger$ \\
        &     & (days)       & (days)   &  (days) & \\[0.5ex]
\hline\\[-1ex]
 \hspace{-0.0mm}$-$2509.0  & I   &  51582.783271  &  0.000640  &   0.000001  & 1 \\
 33946.0  & I   &  55069.581023  &  0.000062  &   0.000044  & 2 \\
 33977.0  & I   &  55072.546051  &  0.000062  &   0.000025  & 2 \\
 40230.0  & I   &  55670.625551  &  0.000028  &   0.000014  & 3 \\
 40251.0  & I   &  55672.634181  &  0.000025  &   0.000059  & 3 \\
 40272.0  & I   &  55674.642702  &  0.000025  &  \hspace{-2.0mm}$-$0.000004  & 3 \\ 
 40272.5  & II  &  55674.690594  &  0.000044  &   0.000065  & 3 \\   
 40273.0  & I   &  55674.738359  &  0.000028  &   0.000006  & 3 \\ 
 40355.5  & II  &  55682.629077  &  0.000123  &  \hspace{-2.0mm}$-$0.000146  & 3 \\ 
 40356.0  & I   &  55682.677091  &  0.000025  &   0.000045  & 3 \\ 
 40429.0  & I   &  55689.659259  &  0.000025  &  \hspace{-2.0mm}$-$0.000011  & 3 \\ 
 41488.0  & I   &  55790.949290  &  0.000025  &  \hspace{-2.0mm}$-$0.000022  & 3 \\ 
 41519.0  & I   &  55793.914357  &  0.000025  &  \hspace{-2.0mm}$-$0.000008  & 3 \\ 
 41529.5  & II  &  55794.918598  &  0.000044  &  \hspace{-2.0mm}$-$0.000058  & 3 \\ 
 41540.0  & I   &  55795.922901  &  0.000026  &  \hspace{-2.0mm}$-$0.000047  & 3 \\ 
 41624.0  & I   &  55803.957263  &  0.000025  &   \hspace{-2.0mm}$-$0.000020  & 3 \\
 41683.0  & I   &  55809.600444  &  0.000041  &  \hspace{-2.0mm}$-$0.000002 & 4\\
 41903.0  & I   &  55830.642731  &  0.000025  &  \hspace{-2.0mm}$-$0.000016 & 2 \\
 42250.0  & I   &  55863.832200  &  0.000024  &  0.000013 & 3\\
 42251.0  & I   &  55863.927831  &  0.000024  &  \hspace{-2.0mm}$-$0.000003 & 3\\
 42461.0  & I   &  55884.013679  &  0.000025  &  0.000021 & 3\\
 42596.0  & I   &  55896.925988  &  0.000024  &  0.000017 & 3 \\[0.5ex]
\hline\\[-2ex]
\end{tabular}
\label{tab:data2}

$^\dagger$ (1) NSVS, (2) Ond\v{r}ejov Observatory, (3) MONET/North,
(4)~Kolonica Observatory,
\end{flushleft}
\vspace{-4mm}
\end{table}

We present here the results of an ongoing search for eclipse time
variations in these two sources, which is based on new observations
extending throughout 2011 and supplemented by archival data reaching
back to 1999. Part of the 2011 observations were performed within the
project {\it PlanetFinders} conducted by high-school teachers and
their students in collaboration with professional astronomers.  This
project provides high school students with first-hand experience in
all aspects of an authentic research project.

\begin{figure*}[t]
\includegraphics[bb=101 45 551 771,height=92mm,angle=-90,clip]{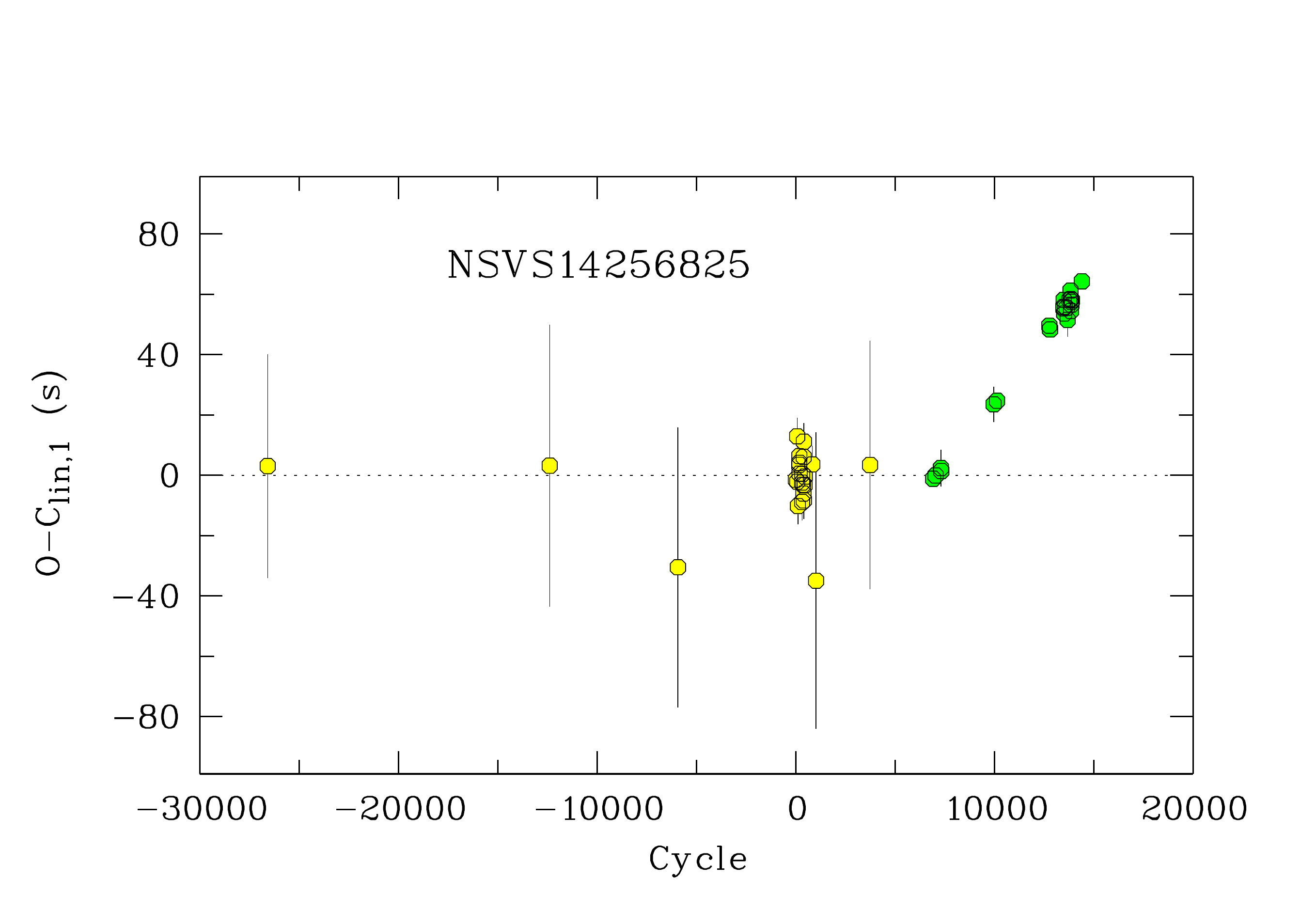}
\hfill
\hspace{1mm}
\includegraphics[bb=101 45 551 740,height=88mm,angle=-90,clip]{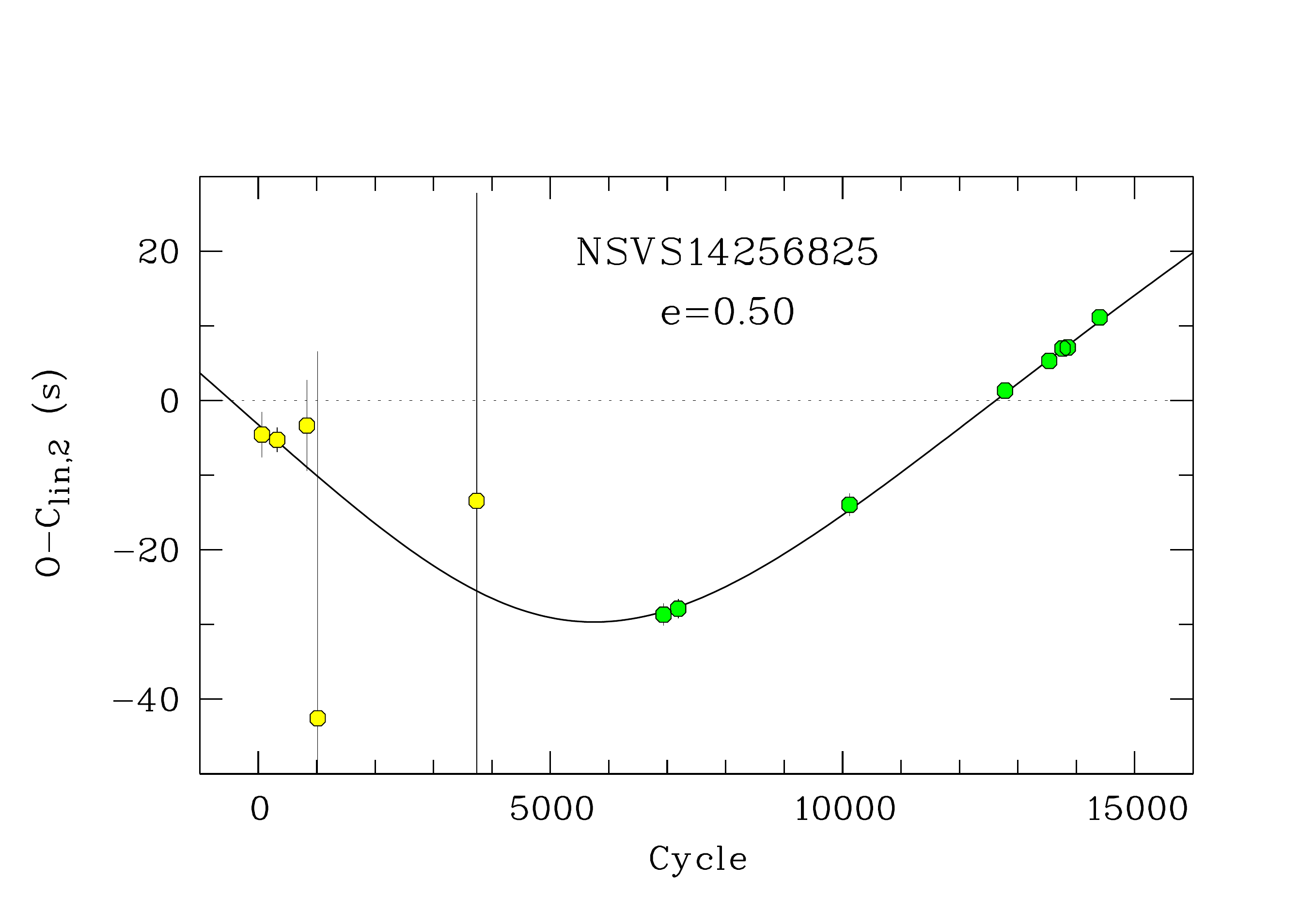}
\caption[chart] {$O-C$ diagrams for NSVS14256825 with previously
  published and our new data shown as yellow and green dots,
  respectively. \emph{Left: } Residuals $O\!-\!C_\mathrm{lin,1}$ for a
  linear ephemeris fitted to the pre-2010 data, cycle numbers $E<9000$
  (dotted line). \emph{Right: } Residuals $O\!-\!C_\mathrm{lin,2}$ for
  an elliptic orbit fit relative to the underlying linear ephemeris of
  the binary (Eq.~4). Weighted mean data points for cyles $E\!\ge\!0$
  are displayed (see text).  }
\label{fig:fig2}
\end{figure*}

\nsvs\ and \hs\ were observed with the MONET/North telescope at the
University of Texas' McDonald Observatory via the MONET browser-based
remote-observing interface between April and December 2011. The
photometric data were taken with an Apogee ALTA E47+ 1k$\times$1k CCD
camera in white light, with exposure times of usually 20\,s. Using
MONET/North, high school students carried out part of the observations
from their classroom.  Additional observations with integration times
between 20\,s and 100\,s were performed with the 1.0-m Vihorlat
National Telescope at Kolonica Saddle, Slovakia, equipped with an FLI
PL1001E CCD in white light; with the 84-cm reflector of the
Observatorio Astronomico Nacional en San Pedro Martir, Mexico, in
Johnson $V$; and with the 65-cm reflector at Ond\v{r}ejov Observatory,
Czech Republic, with a G2-3200 CCD Camera and $R_\mathrm{c}$
filter. All data were subjected to the usual dark current subtraction
and flatfielding. Relative photometry was performed using suitable
comparison stars.  Fig.~1 (left panel) shows a light curve of
\nsvs\ taken with the MONET/N telescope on 5 May 2011 along with the
model fit described below.

Early coverage of \hs\ and \nsvs\ is available from archival data of
the Northern Sky Variability Survey
\citep{wozniaketal04}\footnote{http://skydot.lanl.gov/}, which
provides short glimpses of the sky for an extended period of time in
1999, with exposure times between 20 and 80\,s.  \hs\ was extensively
studied from October 2000 on by \citet{drechseletal01}. The source was
subsequently monitored by \citet[][and references
therein]{qianetal09,qianetal10a}. For \nsvs, on the other hand,
there is a dearth of data between 1999 and 2007, when
\citet{wilsetal07} noted its eclipsing nature. Fortunately, the source
was covered by the All Sky Automated Survey~3 performed from Chile
\citep{pojmanski04}\footnote{http://www.astrouw.edu.pl/~$\sim$asas3/}
with eclipses contained in data from 2003 to 2009. Again, the survey
provides only glimpses of the source with its short series of usually
180\,s exposures in the V-band. By folding the barycentrically and
leap-second corrected data over the known period, mean eclipse light
curves can be reconstructed from both the NSVS and the ASAS3
data. These are displayed in the center and right panel of Fig.~1. The
model light curves employ parameters fixed at the values determined
from fits to the MONET/N data (see next paragraph). The NSVS data that
contain eclipses of \nsvs\ cover the period from 19 April to 10 August
1999. The ASAS3 data can be broken up into four light curves
defining representative mid-eclipse times for dates in 2003, 2005,
2007, and 2008.

We fitted the eclipse light curves obtained with the MONET/N, Kolonica,
Ond\v{r}ejov, and San Pedro Martir telescopes, using a heuristic model
that involves no physics, except requiring symmetry about the
mid-eclipse time $p_1$. In the variable $\tau\!=\!t-p_1$, with $t$ the
time, the observed relative flux $F(t)$ is represented by a modified
and truncated inverted Gaussian $G(\tau)$ multiplied by a
polynomial $P(\tau)$ that accounts for any real or apparent variation
in the out-of-eclipse flux, i.e.,
\begin{eqnarray}
F(t) & = & P(\tau)\,\mathrm{min}\left(1,G(\tau)\right), \\
\mathrm{\noindent where} &&\nonumber \\
G(\tau) & = & p_2 - p_3\,\mathrm{exp}\left [-\frac{1}{2}\left( \frac{\mid\!\tau\!\mid}{p_4}\right)^{p_5}\right ], \\  
P(\tau) & = & p_6 +p_7\,\tau +p_8\,\tau^2, 
\end{eqnarray}
and $p_\mathrm{i}, ~\mathrm{i}\!=\!1\ldots8$, are the fit parameters. The
Gaussian is modified by replacing the square in the exponential with a
free parameter $p_5$, which allows the creation of a more peaked or
broader shape of the eclipse light curve. With a zero level $p_2$
exceeding unity, the truncation provided by the prescription of Eq.\,1
reproduces the finite base width of the eclipse. At each step of the
iterative fitting procedure, $F(t)$ is furthermore averaged over the
finite, not necessarily equidistant exposure intervals.

Fig.~1 shows that a close to perfect fit to the observed light curve
is achieved. For exposure bins of 20\,s, the achievable accuracy of
the derived mid-eclipse times is $\sim\!1$\,s for the primary eclipses
and a few seconds for the secondary eclipses.  The formal errors in
the mid-eclipse times derived from fits to the NSVS and ASAS3 data
amount to about 1 minute.

\section{Results}

\subsection{NSVS14256825}

Table~1 lists the new mid-eclipse times for \nsvs, along with the
formal errors from the light curve fits. These data complement the
only previously published ones from \citet{wilsetal07}. The first five
entries are from the NSVS for 1999 and the ASAS3 for time intervals of
a few months in the summers of 2003, 2005, 2007, and 2008. These
eclipse times are obtained from fits of the known eclipse profile to
the barycentrically corrected data folded over the orbital period. The
quoted Julian day is that of the lowest data point during the eclipse,
but the fraction of the day is derived from the fit. Of the remaining
27 entries, 24 represent primary eclipses (I) and three secondary
eclipses (II). All times were corrected to the Solar System barycenter
with leap seconds added and quoted as BJD(TT). We derived an ephemeris
by adding the mid-eclipse times of \citet{wilsetal07} converted from
HJD to
BJD(TT)\footnote{http://astroutils.astronomy.ohio-state.edu/time/}.

Using either the data of \citet{wilsetal07} alone or all primary
mid-eclipse times until September 2009 (cycle number 7322), we obtained
the same orbital periods within the errors, $P\!=\!0.110374086(88)$\,days
and $P\!=\!0.110374093(3)$\,days, respectively. There is no evidence
of a period change in the pre-2010 data. Fig.~2 (left panel) shows
the $O\!-\!C_\mathrm{lin,1}$ values relative to a linear ephemeris
with the latter period. The situation changes drastically for the
post-2009 data, where a sharp upturn is observed, which corresponds to a
rather abrupt period increase by 9\,ms. A physical mechanism that can
cause such a change in a detached binary is its response to a
companion in a highly elliptic orbit hurdling through
periastron. Similar variations in QS\,Vir \citep{parsonsetal10} and
DP\,Leo \citep{beuermannetal11} have been ascribed to the action of
third bodies. While such an interpretation may seem premature for
\nsvs, it fits the available data exceedingly well and provides a
definite prediction, which can be tested in the next couple of years.

Given the uncertain orbital period of the putative third body, the
eccentricity is likewise uncertain, a situation that is similarly met
for some comets in the Solar System observed only near perihelion.
Fixing the period at $P_3\!=\! 20$ yr, similar to values found for
other PCEBs, we fitted all data with the light-travel time effect produced by
a third body in an elliptic orbit.  The fit involves seven free
parameters, five for the elliptic orbit and two for the underlying
linear ephemeris of the binary
\citep{beuermannetal10,beuermannetal11}. The latter is
\begin{equation}
\mathrm{BJD(TT)} = 54274.208923(4) + 0.1103741324(3)\,E,
\end{equation}        
where $E$ is the cycle number. The best-fit eccentricity of the
20-year orbit is $e_3\!=\!0.50$, but a valley of low \,$\chi^2$ in the
$P_3, e_3$ plane stretches from 14\,yr, 0.30 to beyond
30\,yr, 0.70. While we fitted all individual mid-eclipse times, a
clearer graphical presentation is obtained by plotting the weighted
mean $O\!-\!C_\mathrm{lin,2}$ values relative to the ephemeris of
Eq.~4 for subgroups of data taken close together in time (Fig.~2,
right panel). The subgroups combine data taken over intervals of a
week to about a month, which is much shorter than the suggested orbital
period. The residuals of the individual new timings from the elliptic
orbit fit are quoted in Table~1. The amplitude of the
$O\!-\!C_\mathrm{lin,2}$ variation, $K_3\!\simeq\! 59$\,s, suggests a
mass of the third body of $M_3\!\simeq\! 12$\,\mjup, if we adopt a
total mass of the binary of 0.62\,\msun, which is similar to those of
\hs\ \citep[0.617\,\msun, ][]{drechseletal01} and HW~Vir
\citep[0.627\,\msun, ][]{leeetal09}. The value of $M_3$ varies with
the uncertain orbital period $P_3$ as $P_3^{-2/3}$ and increases with
the unknown inclination $i_3$ of the third body as 1/sin\,$i_3$. If
the orbit of the companion is roughly co-planar with the binary,
it qualifies as a giant planet.  For the 20-year orbit, the semi-major
axis is 6.3\,AU. Our fit yields an argument of periastron for the
motion of the center of gravity of the binary measured from the
ascending node in the plane of the sky
$\varpi_\mathrm{bin}\!=\varpi_3-180^\circ\!\!\simeq\!-98^\circ$. Periastron
passage occurred near JD\,2454836 ($E\!\simeq\!  5091$).  The fit
suggests that $O\!-\!C_\mathrm{lin,2}$ will reach a maximum of 88\,s
in 2019 ($E\!\simeq\!40000$), with actual numbers depending on the
uncertain value of $P_3$.

\subsection{HS0705+6700}

A large body of mid-eclipse times of \hs\ has been collected by
\citet{qianetal09,qianetal10a} and interpreted in terms of a third
body with a mass of about 30\,\mjup, orbiting the binary with a period
of 7.15\,yr.  \citet{camurdanetal12} added a few timings and found a
similar period. We obtained 18 more primary and three secondary
eclipses of \hs, significantly extending the coverage. In addition, we
recovered an early eclipse timing from the NSVS. The new data are
listed in Table~2.

\begin{figure}[t]
\includegraphics[bb=101 33 551 701,height=89mm,angle=-90,clip]{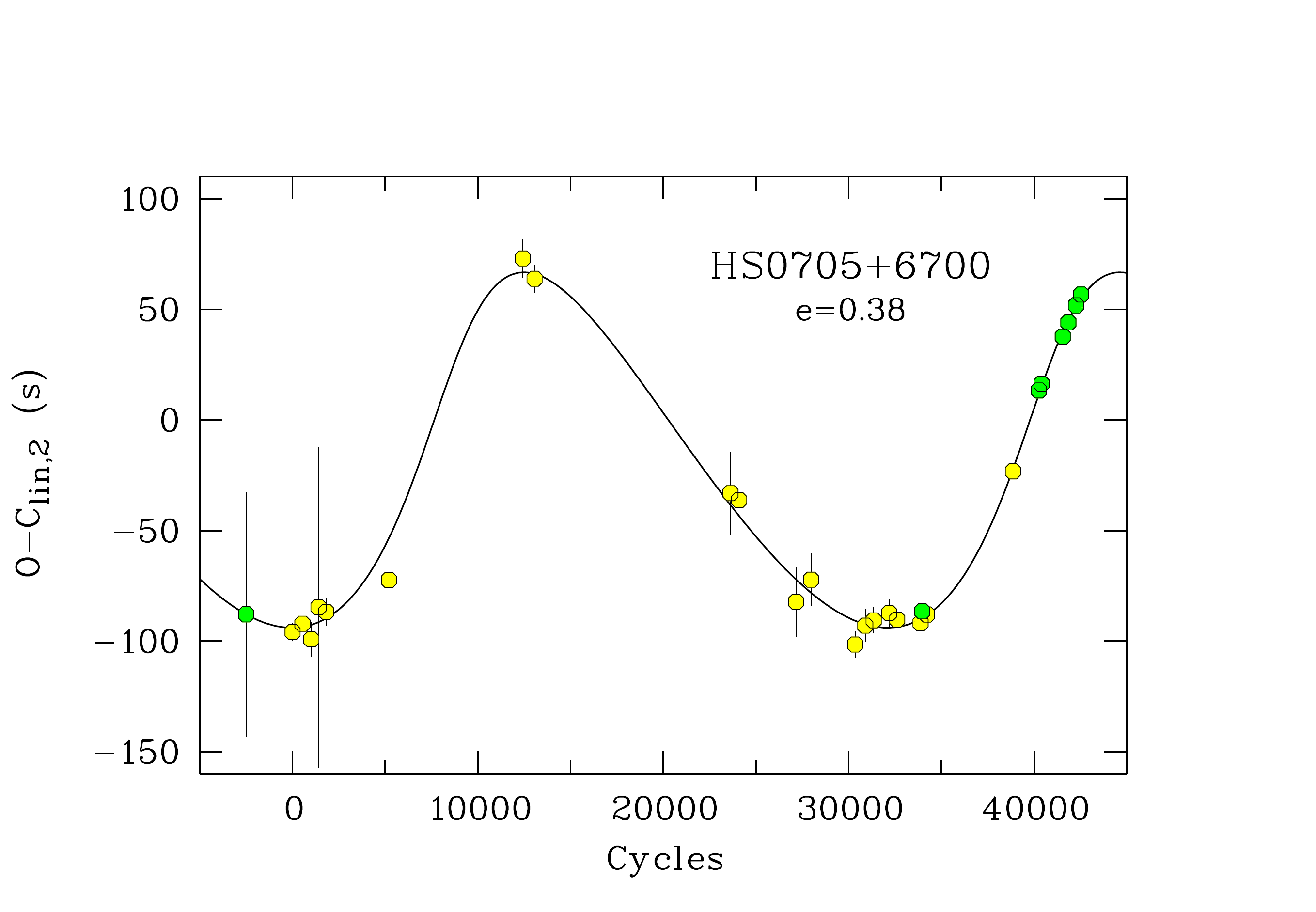}
\caption[chart]{ Elliptic orbit fit for HS\,0705+6700. Shown are the
  residuals $O\!-\!C_\mathrm{lin,2}$ relative to the underlying linear
  ephemeris of the binary (Eq.\,5).  Color coding is as in Fig.~2.}
\label{fig:fig3}
\end{figure}

We fitted the entire set of 89 primary mid-eclipse times assuming that
the variations are caused by a third body in an elliptic orbit. The
fit yields \,$\chi^2\!=\! 80.6$ for 82 degrees of freedom and is shown
in Fig.~3. For a clearer presentation, we collected the individual
data points again into appropriate subgroups comprising time intervals
of up to a week for the more accurate data and one month for the less
accurate ones near minimum $O\!-\!C_\mathrm{lin,2}$ (compare Fig.~3 of
Qian et al. 2010a, where the individual $O-C$ values are displayed).
The underlying intrinsic ephemeris of the binary is
\begin{equation}
\mathrm{BJD(TT)} = 51822.761677(6) + 0.0956466253(2)\,E.
\end{equation}
The amplitude of the light-time effect produced by the third body is
$K_3\!=\! 86\pm1$\,s, its orbital period is $P_3\!=\!
8.41\pm0.05$\,yr, and the eccentricity is $e_3= 0.38\pm0.05$. The
argument of periastron is
$\varpi_\mathrm{bin}\!=\varpi_3-180^\circ\,\!\!=\!26^\circ\pm2^\circ$
and periastron passage occurred near JD\, 2455719 ($E\simeq 40738$) at
$O\!-\!C\!\simeq\!22$\,s. For the present model, a maximum of
$O\!-\!C\!\simeq\!67$\,s is predicted to occur near $E\!\simeq\!44700$
in June 2012. With a binary mass of
0.617\,\msun\ \citep{drechseletal01}, the semi-major axis of the third
object is 3.52\,AU and its mass is
$M_3\!=\!(31.5\pm1.0)/\mathrm{sin}\,i_3$\,\mjup, very similar to the
values quoted by \citet{qianetal10a}. This mass is significantly lower
than the estimate of \citet{camurdanetal12}. The present data provide
no evidence of a deviation of the mid-eclipse times from those created
by a single object, but the achievable accuracy of better than 1\,s
opens the prospect to searching for a fourth object, as found, e.g.,
in the WD/MS binary NN~Ser \citep{beuermannetal10}.

\subsection{Secondary eclipses}

The present observations locate the secondary eclipse in \nsvs\ and
\hs\ at orbital phases of $\phi\!=\!0.5002\pm0.0002$ and
$\phi\!=\!0.4997\pm0.0003$, respectively. This finding along with
the clearly non-sinusoidal form of the eclipse time variations
excludes a sizeable contribution of apsidal motion \citep{todoran72} to the
observed effect.

\section{Discussion}

We have presented evidence of eclipse time variations in \nsvs\ and
\hs, which are closely in line with variations expected from the
light-travel time effect of a third body. The existence of companions
orbiting PCEBs can no longer be easily refuted, although the cases of
UX~UMa and perhaps HU~Aqr illustrate that there should be no
indiscriminate application of the planetary model to all PCEBs showing
eclipse time variations. In \hs, the relatively short period of the
third body of 8.4 yr and the large amplitude of 84\,s will allow us to
test the periodicity of the signal within the next decade, and
confirmation will provide support for the third-body model. The often
considered alternative of spin--orbit coupling, where star cycles
change the spin of the secondary star and thereby the angular
momentum of the binary orbit \citep{applegate92}, remains a contender,
but may be too feeble to account for the observed effects
\citep[e.g.][]{brinkworthetal06,chen09,schwarzetal09,potteretal11}.
Apsidal motion, can be safely excluded, at least for the objects
studied here.

Given the large number of planets detected around single stars, it is
unlikely that substellar objects do not exist in orbits around close
binaries. The observational problem is to differentiate between the
$O\!-\!C$ variations caused by these companions and any effects
intrinsic to the binary, which would require data covering many
years. The fundamental questions then relate to the origin of such
companions, which need not be the same for planet-sized and
brown-dwarf objects, and to the incidence in different subtypes of
PCEBs. With HW\,Vir \citep{leeetal09}, \hs, and probably \nsvs, at
least three PCEBs with sdB primaries possess companions. For systems
with a WD primary, the companions to NN~Ser \citep{beuermannetal10}
and DP\,Leo \citep{qianetal10b,beuermannetal11} appear reasonably
well-established. Depending on the type of primary, these systems have
different evolutionary histories \citep{zorotovicetal11}. For the
former, the mass of the sdB star indicates that they are in the
He-burning stage, having formed on the first giant branch. On the
other hand, NN~Ser and DP~Leo contain CO-WDs and have formed on the
asymptotic giant branch. The mass distribution of white dwarfs from
the SDSS \citep{rebassaetal11}\footnote{http://www.sdss-wdms.org}
suggests that some PCEBs contain He-WDs and it will be interesting to
find out whether these systems, too, contain companions, or more
generally planetary systems, a finding that may shed some light on the
origin of companions to PCEBs.

\begin{acknowledgements}
This work is based in part on data obtained with the MOnitoring
NEtwork of Telescopes (MONET), funded by the Alfried Krupp von Bohlen
und Halbach Foundation, Essen, and operated by the
Georg-August-Universit\"at G\"ottingen, the McDonald Observatory of
the University of Texas at Austin, and the South African Astronomical
Observatory.  The "Astronomie \& Internet" program of the Foundation
and the MONET consortium provides a major fraction of the observation
time to astronomical projects in high schools worldwide. The work of
JD and his students at the Max-Planck-Gymnasium was supported by the
Robert Bosch Foundation. The research of MW and PZ was supported by
the Research Program MSM0021620860 of the Ministry of Education of the
Czech Republic.

\end{acknowledgements}

\bibliographystyle{aa}

\end{document}